\documentclass[a4paper,fleqn,usenatbib]{mnras}

\usepackage{newtxtext,newtxmath}

\usepackage[T1]{fontenc}
\usepackage{ae,aecompl}

\usepackage{graphicx}   
\usepackage{amsmath}    
\usepackage{amssymb}    

\title[Short title, max. 45 characters]{Energy accumulation mechanism in pulsar magnetospheric plasma eigen-waves and formation of Giant Radio Pulses}

\author[G. Machabeli et al.]{
G. Machabeli,$^{1}$\thanks{E-mail: g.machabeli@iliauni.edu.ge}, N.
Chkheidze$^{1}$ and I. Malov$^{2}$
\\
$^{1}$Centre for Theoretical Astrophysics, ITP, Ilia State
University, Tbilisi 0162, Georgia\\
$^{2}$P.N.Lebedev Physical Institute of the Russian Academy of
Sciences, 53 Leninskiy Prospect, Moscow, 119991, Russia\\}

\pubyear{2018}

\begin{document}
\label{firstpage}
\pagerange{\pageref{firstpage}--\pageref{lastpage}} \maketitle

\begin{abstract}
In the present work we consider energy accumulation mechanism in
relativistic electron-positron plasma in the magnetosphere of
pulsars. Waves propagating almost across the magnetic field lines
are generated that accumulate the energy of plasma particles, as the
waves stay significantly long in the resonance region. The
accumulated energy is transmitted in the parallel direction of the
magnetic field lines as soon as the non-linear plasma processes
start to operate. It is suggested that exit of the accumulated
energy via waves propagating along the magnetic field lines matching
the viewing angle of the observer produces giant pulses. It is shown
that these waves come in the radio domain, explaining the Giant
Radio Pulse phenomenon.
\end{abstract}

\begin{keywords}
keyword1 -- keyword2 -- keyword3
\end{keywords}

\section{Introduction}

From several pulsars sporadic intense radio pulses are observed
known as the Giant Radio Pulses (hereafter GRPs) that are much
brighter than the regular pulses. This rare phenomenon has been
detected only in 16 pulsars (see parameters in \citet{kaza}), one of
the first of them was the Crab pulsar PSR B0531+21 \citep{stalin,
argyle}. The peak flux densities of GRPs can exceed hundreds and
thousand of times the peak flux density of regular pulses and the
pulses reveal extremely narrow width in comparison with the average
emission of the pulsar (their duration is of the order of several
microseconds down to few nanoseconds \citet{istomin,eilek}). The
ultrashort durations of the giant pulses imply very high equivalent
brightness temperatures \citep{hankins} indicating that they
originate from nonthermal coherent emission processes. Moreover,
through the GRPs narrow-band radiation is emitted (the width of the
spectrum is of the order of the frequency band; \citet{popov}).
Another distinctive feature of giant pulses from usual radio
emission of pulsars is that their amplitude distribution is power
law, while that of normal pulses follows a log-normal distribution
\citep{argyle}. All of the listed features of giant pulses that
differs them from the regular radio pulses indicates a different
emission generation scenario.

Knowing the origin of giant radio pulses is extremely important for
understanding the difference between "normal" radio pulsars and the
ones emitting also the giant pulses. Although, several theoretical
models explaining the GRPs have been proposed, the emission
mechanism of giant pulses still remains unclear. \citet{weatherall},
suggested that strong electrostatic turbulence in electron-positron
($e^{-}e^{+}$) plasma could produce intense radiation.
\citet{hankins} claimed that GRPs from the Crab pulsar is produced
through the conversion of electrostatic turbulence in the pulsar
magnetosphere by the mechanism of spatial collapse of nonlinear wave
packets. In \citet{petrova} was proposed that GRPs are generated by
induced scattering of low frequency radiation in the pulsar
magnetosphere causing a redistribution of the radio emission in
frequency. It has been found that the efficiency of the
amplification of radiation drastically depends on radio luminosity
of a pulsar. For the pulsars with GRPs, PSR B0531+21 and PSR
B1937+21 the value of total radio luminosity
$L_{r}\simeq4\cdot10^{31}$erg/s and $7.44\cdot10^{30}$erg/s,
consequently. These values are indeed higher than that for other
"normal" radio pulsars \citet{malov}. In \citet{istomin} it was
suggested that the radio emission generation process is implemented
close to the light cylinder region by the electric discharge taking
place due to the magnetic reconnection of the field lines connecting
the opposite magnetic poles. It is considered that the GRPs in
pulsars PSR B0531+21 and B1937+21 are generated through the maser
amplification of Alfv\'{e}n waves. It is supposed that these objects
are perpendicular rotators which provides specific charge
distribution on magnetic poles of the star. This causes appearance
of strong electric field and magnetic reconnection close to the
light cylinder that induces maser amplification of the amplitude of
Alfv\'{e}n waves. The mentioned process can develop only if the
value of the magnetic field in the region of the light cylinder is
sufficiently high (of the order of $10^{6}$G,that is well fulfilled
for the considered pulsars). The most recent explanation was
provided by \citet{lyutikov} that is closest to our scenario
suggested in the present work. He supposed that GRPs are generated
on closed magnetic field lines in a limited small volume, where the
last closed field lines approach the light cylinder via anomalous
cyclotron resonance on the ordinary mode. This model is especially
developed for Crab pulsars GRPs associated with the interpulse.

Here we present an alternative explanation of the phenomena of GRPs,
which suggests the mechanism of radio emission energy growth due to
nonlinear plasma processes taking place in the pulsar magnetosphere.
Particularly, the GRPs generation scenario implies that the energy
is accumulated in waves propagating practically across the local
magnetic field lines that subsequently change the propagating
direction due to induced scattering of waves on plasma particles and
finally can be observed as pulsar radio emission. This happens when
the wave vector is practically parallel to the direction of magnetic
field lines. In general we believe that the pulsed radio emission
from pulsars is generated near the light cylinder region by plasma
instabilities developing in the outflowing plasma on the open field
lines of the pulsar magnetosphere. Plasma can be considered as an
active medium that amplifies its normal modes through the resonant
wave-particle interaction. The generated $e^{-}e^{+}$ plasma eigen
waves propagate along the magnetic field lines and are in this case
vacuum-like electromagnetic waves so they can leave the
magnetosphere directly and reach an observer as pulsar radio
emission. This plasma emission model has been well developed and
explains all main observational features of radio pulsars (see
\citet{lo,machus,ka_a,ly,lu}).

\begin{figure}
      \centering {\includegraphics[width=0.35 \textwidth]{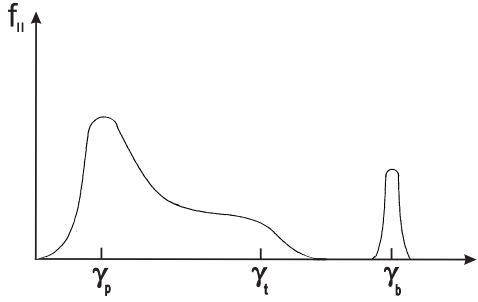}}
    \caption{Distribution function for a one-dimensional electron-positron
plasma of pulsar magnetosphere.}
    \label{fig1}
\end{figure}

According to our scenario the formation of GRPs in pulsar
magnetosphere is based on the above mentioned radio emission
generation model. Therefore, it is foremost essential to introduce
the latter in more details. Consequently, in Section 2, we describe
the radio emission generation mechanism. This implies considering
the maser-type plasma instabilities, anomalous cyclotron-Cherenkov
and Cherenkov-drift resonances generating waves propagating along
(Section 2) and across (Section 3) the pulsar's magnetic field
lines. The waves propagating almost across the magnetic field
accumulate energy in waves until the nonlinear processes start to
operate. In Section 4 the development of nonlinear processes is
considered that plays the main role in producing the GRPs.

\section{Generation of waves propagating along the magnetic field - Radio emission model}

\begin{figure}
      \centering {\includegraphics[width=0.35 \textwidth]{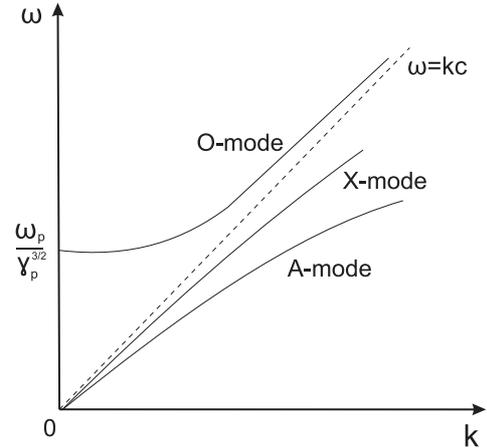}}
    \caption{Dispersion curves for the eigen-waves in a magnetized
    electron-positron plasma for oblique propagation \citep{vladi}.}
    \label{fig2}
\end{figure}

According to the works of \citet{st} and \citet{tade}, due to the
cascade processes of pair creation, a pulsar's magnetosphere is
filled by $e^{-}e^{+}$ plasma with an anisotropic one-dimensional
distribution function (see Figure 1) and consists of three
components. First component is the most energetic primary beam,
which consists of primary electrons (the maximum Lorentz factor of
the particles $\gamma_{b}\sim10^{7}$ ) extracted from the star
surface by electric field induced due to rotation of the magnetized
neutron star. According to Goldreich-Julian the density of the
primary beam $n_{b}=n_{GJ}=7\cdot10^{-2}B_{0}P^{-1}$, where $B_{0}$
is the magnetic field at pulsar surface and $P$ is the spin period.
The second component of pulsar magnetospheric plasma is the bulk of
plasma that consists of secondary electrons and positrons produced
through the cascade process of pair creation taking place near the
polar cap in the vacuum gap region \citep{mich,rude}. The typical
Lorentz factor of the plasma particles assuming that the pulsar has
a dipole magnetic field is $\gamma_{p}\approx10-10^{3}$ and for the
quadrupole model for the pulsar magnetic field
$\gamma_{p}\approx2-5$ \citep{machus}. Considering the quadrupole
model for the Crab pulsar, its parameters $B_{0}=7.6\cdot 10^{12}$G,
$P=33$ms and the equipartition of energy among the plasma components
$2n_{p}\gamma_{p}\approx n_{b}\gamma_{b}$ one can obtain
$n_{p}\approx 10^{20}$cm$^{-3}$. The third component is a tail on
the distribution function with Lorentz factor
$\gamma_{t}\sim10^{4-5}$. Plasma with anisotropic distribution
function is unstable and can cause excitation of plasma eigen-waves.
The properties of the $e^{-}e^{+}$ plasma have been investigated
quite thoroughly (e.g., \citet{vo,aro,lo1}). There are three
fundamental modes: the transverse extraordinary (X-mode) wave with
the electric field perpendicular to the $\mathbf{k}$ and
$\mathbf{B}$ (where $\mathbf{k}$ is the wavevector and $\mathbf{B}$
is the vector of the magnetic field) and the longitudinal-transverse
modes with the electric field lying in the plane formed by
$\mathbf{k}$ and $\mathbf{B}$: the ordinary (O-mode) and Alfv\'{e}n
(A-mode) modes. The O mode on the diagram $\omega(k)$ begins with
the Langmuir frequency (Figure 2) and when $k_{\perp}= 0$, it
reduces to the pure longitudinal Langmuir wave. In the laboratory
frame the dispersion relations of O, A and X-modes are,
respectively:
\begin{eqnarray}\label{}
    \omega_{t}=kc(1-\delta), \qquad \qquad \textrm{X-mode} \\
    \omega_{lt}=k_{\parallel}c\left(1-\delta-\frac{k_{\perp}^{2}c^{2}}{4\gamma_{p}\omega_{p}^{2}}\right), \qquad \qquad
    \textrm{A-mode} \\
    \omega^{2}_{Lt}=\omega_{p}^{2}\gamma_{p}^{-3}+k^{2}c^{2}, \qquad \qquad \textrm{O-mode}
\end{eqnarray}
where $\delta=\omega_{p}^{2}/4\gamma_{p}^{3}\omega_{B}^{2}$ and
$\omega_{p}^{2}=4\pi n _{p}e^{2}/m$. The indexes p and b correspond
to the bulk of plasma and the primary beam of electrons.

The names of the waves used above (X, A, O modes) are often used by
analogy with names for the waves in nonrelativistic electron-ion
plasma. However, the waves in relativistic pair plasma are quite
different. The low-frequency part of the X-mode is in the
superluminal region, where $\upsilon_{ph}>c$ and its generation is
only possible in the region where it intersects the line
$\omega_{t}=kc$ on the $\omega(k)$ diagram (See Fig. x). In the
high-frequency region $\omega\gg\omega_{p}/\gamma_{p}^{3/2}$ the
O-mode merges with the X-mode and converts into transverse wave and
for the region where $\omega<\omega_{p}/\gamma_{p}^{3/2}$ the O-mode
describes electrostatic Langmuir waves, the frequency of which at
the intersection point with the line $\omega_{t}=kc$ equals to
$\omega^{2}=2\omega_{p}^{2}\gamma_{p}$ \citep{mixa}. For more
clarity we prefer here to use different designations for the X, A,
and O-modes, correspondingly naming them as $t$, $lt$ and $Lt$
waves. The magnetospheric plasma is anisotropic that is quite
natural in the presence of strong pulsar magnetic field. The
relativistic particles efficiently emit through the synchrotron
mechanism, quickly losing transverse momenta near the star surface
continuing to move along the magnetic field lines with the
relativistic momenta $p_{\parallel}=m\upsilon_{\parallel}\gamma \gg
p_{\perp}$. Such plasma is unstable causing excitation of
electromagnetic waves and as a result both types of eigen-waves $t$,
as well as $lt$ can be generated \citep{lo}. When $t$ and $lt$ waves
are generated with the small inclination angle with respect to the
magnetic field, they propagate in the same direction along the field
lines and are orthogonally polarized. These are vacuum-like
electromagnetic waves and can leave the magnetosphere reaching an
observer as pulsar emission.

When considering generation of the radio waves, for which the
wavelength $\lambda$ is much bigger than the average distance
between the plasma particles $\lambda \gg n^{-1/3}$ (here $n$ is the
density of plasma particles) the effects of wave interference should
be taken into account. As already mentioned the distribution
function shown on Fig. x tends to be unstable relative to some
plasma instabilities. The strongest instabilities that can develop
in the pulsar magnetosphere are the Cherenkov-drift and anomalous
cyclotron-Cherenkov instabilities. The frequency of the waves
generated via the mentioned resonances for typical pulsars falls
within the radio band. The Cherenkov-drift instability develops at
the resonance:
\begin{equation}\label{}
    \omega-k_{\varphi}\upsilon_{\varphi}-k_{x}u_{x}=0,
\end{equation}
where $u_{x}=\upsilon_{\varphi}^{2}\gamma_{res}/\omega_{B}R_{B}$ is
the drift velocity of the particles across the magnetic field caused
by the magnetic field inhomogeneity and directed along the direction
of the x-axis, $\gamma_{res}$ is the Lorentz factor of the resonant
particles, $\omega_{B}=eB/mc$ is the cyclotron frequency and $R_{c}$
is the curvature radius of the magnetic field lines. Here the
cylindrical coordinates $x$, $r$, $\varphi$ have been chosen,
$x$-axis directed transversely to the plane of the curved field
line, with $r$ directed along the radius of curvature of the field
line and $\varphi$ the azimuthal coordinate; $k_{\varphi}$ is the
component of the wavevector along the magnetic field and
$k_{\perp}=(k_{r}^{2}+k_{x}^{2})^{1/2}$.

The second strongest instability in the pulsar magnetosphere the
cyclotron-Cherenkov instability, which generates low frequency waves
via the anomalous Doppler effect develops at the resonance:
\begin{equation}\label{}
    \omega-k_{\varphi}\upsilon_{\varphi}-k_{x}u_{x}+\frac{\omega_{B}}{\gamma_{res}}=0.
\end{equation}
Lets us consider this resonance condition in more details. Given
that $\upsilon_{\varphi}\approx
c(1-1/2\gamma_{res}^{2}-u_{x}^{2}/c^{2})$,
$k=k_{\varphi}(1+k_{\perp}^{2}/k_{\varphi}^{2})^{1/2}\approx
k_{\varphi}(1+k_{\perp}^{2}/2k_{\varphi}^{2})$ for the $t$ waves
with the spectrum (1) and small angles of propagation one can
rewrite the resonance condition (5) in the following way
\begin{equation}\label{}
     \frac{1}{2\gamma_{res}^{2}}+\frac{1}{2}\left(\frac{k_{x}}{k_{\varphi}}-
     \frac{u_{x}}{c}\right)^{2}+\frac{k_{r}^{2}}{2k_{\varphi}^{2}}-\delta=
     -\frac{\omega_{B}}{\gamma_{res}k_{\varphi}c}.
\end{equation}
For the typical pulsar parameters the conditions (6) can be
fulfilled for both the particles in the tail of the distribution
function and the primary beam electrons. Although, the
Cherenkov-drift resonance condition (4) is exclusively fulfilled for
the beam particles. For the bulk plasma $\gamma_{res}=\gamma_{p}$
the resonance can not be implemented as in this case the term
$1/\gamma_{p}^{2}$ is the largest.

For the wave generation via the cyclotron resonance as it follows
from expression (6) it is clear that the following inequality should
be fulfilled:
\begin{equation}\label{}
    \delta>\frac{1}{2\gamma_{res}^{2}}+\frac{1}{2}\left(\frac{k_{x}}{k_{\varphi}}-
     \frac{u_{x}}{c}\right)^{2}+\frac{k_{r}^{2}}{2k_{\varphi}^{2}}.
\end{equation}
The parameter $\delta$ is quite small, which makes it challenging
fulfillment of the condition (7). Although, the quantity
$\omega_{p}^{2}/\omega_{B}^{2}\sim B$ and consequently grows with
the growth of $r$ - distance to the pulsar and for typical pulsar
parameters this condition can be fulfilled for the distances of the
order of light cylinder radius.

Let us estimate the frequency of the $t$ waves generated via the
cyclotron resonance. If we denote the angle between the wave vector
$\mathbf{k}$ and the magnetic field $\mathbf{B}$ as $\theta$ and
consider the small propagation angles $\theta\ll1$ and  take into
account that $u_{x}/c\ll1$, after neglecting the drift term one
obtains:
\begin{equation}\label{}
    \frac{1}{2\gamma_{res}^{2}}+\frac{\theta^{2}}{2}-\delta=-\frac{\omega_{B}}{\omega
    \gamma_{res}}.
\end{equation}
The Eq. (8) requires that $1/2\gamma_{res}^{2}<\delta$ and
$\theta^{2}/2<\delta$. The first condition implies that the resonant
particles should be moving with a velocity faster than the phase
velocity of the wave. The second condition limits the generated
emission to small angles with respect to the magnetic field.
Assuming that $1/2\gamma_{res}^{2}\ll\delta$ and
$\theta^{2}/2\ll\delta$ from Eq. (8) one can estimate the frequency
of the generated waves:
\begin{equation}\label{}
    \omega\simeq\frac{\omega_{B}}{\delta \gamma_{res}},
\end{equation}
which comes in the radio frequencies for the typical pulsar
parameters.

Both cyclotron-Cherenkov and Cherenkov-drift instabilities are
capable to explain the main observational characteristics of pulsar
radio emission. The cyclotron-Cherenkov instability is responsible
for the generation of the core-type radio emission and the
Cherenkov-drift instability is responsible for the generation of the
cone-type emission. The excited waves propagate along the magnetic
field lines and the frequency of the generated waves for typical
pulsars falls within the radio band. As shown in \citet{machus1} the
$t$ waves can be only excited by anomalous Doppler resonance,
whereas Cherenkov-drift resonance can generate both $t$ and $lt$
waves. These instabilities occur in the outer parts of the
magnetosphere in the region near the light cylinder. The location of
the emission region is determined by the corresponding resonant
condition for the instabilities. Instabilities develop in a limited
region on the open field lines. The size of the emission region is
determined by the curvature of the magnetic field lines, which
limits the length of the resonant wave-particle interaction. The
location of the cyclotron instability is restricted to those field
lines with large radii of curvature, while the Cherenkov-drift
instability occurs on field lines with curvature bounded both from
above and from below. Thus, both instabilities produce narrow
pulses, although they operate at radii where the opening angle of
the open field lines is large.

\section{Generation of waves propagating across the magnetic field - The drift waves}

Until now we have discussed generation and propagation of waves with
frequencies falling within the radio band and the rather narrow
angles of propagation $\theta\approx k_{\perp}/k_{\varphi}\ll1$.
Taking into account the weak inhomogeneity of the magnetic field
leads to drift of the plasma particles across the field lines
violating the cylindrical symmetry and the permittivity tensor
occurs additional terms that are proportional to the expression
$1/(\omega-k_{\varphi}\upsilon_{\varphi}-k_{x}u_{x})$ assuring
generation of waves perpendicular to the magnetic field. It should
be mentioned that the inhomogeneity of the medium (curvature of the
magnetic field lines of pulsar) is revealed in linear approximation
only in drift of the charged particles. Thus, when $kR_{B}\gg1$ the
waves propagating in $e^{-}e^{+}$ plasma in the pulsar magnetosphere
do not feel the curvature of the field lines. Though, for the
ultrarelativistic particles (e.g. primary beam electrons) one can
not neglect the drift motion across the magnetic field lines when
considering the wave generation processes.

Let us now investigate generation of low-frequency $lt$ waves via
the Cherenkov-drift resonance propagating nearly transversely to the
magnetic field $\theta=\pi/2$. In \citet{ka_b} it is shown that for
waves propagating almost across the magnetic field lines and taking
into account that the parameters $\gamma\omega/\omega_{B}\ll1$,
$(u_{x_{\alpha}}^{2}/c^{2})\ll1$ and $k_{\varphi}/k_{x}\ll1$ are
small, the dispersion relation can be written as
\begin{equation}\label{}
    \frac{k_{x}^{2}c^{2}}{\omega^{2}-k_{\varphi}^{2}c^{2}}=\varepsilon_{\varphi\varphi}=1+\Sigma_{\alpha}\frac{\omega_{p_{\alpha}}^{2}}{\omega}
    \int\frac{\upsilon_{\varphi}/c}{(\omega-k_{x}u_{x_{\alpha}}-k_{\varphi}\upsilon_{\varphi_{\alpha}})}\frac{\partial f}{\partial
    \gamma}dp_{\varphi}.
\end{equation}
The sum is taken over the all particle species, electrons and
positrons of the bulk plasma, tail particles and electrons of the
beam ($\alpha=p,t,b$). Assuming that
\begin{equation}\label{}
    \omega=k_{x}u_{x}+k_{\varphi}\upsilon_{\varphi}+a,
\end{equation}
where $a$ is the small parameter and
$u_{x}=\upsilon_{\varphi}^{2}\gamma_{b}/\omega_{B}R_{B}$ (the
resonant particles are the fastest beam electrons). After
integration the dispersion (10) can be rewritten as:
\begin{equation}\label{}
     1-\frac{3\omega_{p}^{2}}{2\gamma_{p}^{3}\omega^{2}}-\frac{\omega_{b}^{2}}{\omega a^{2}}
     \frac{u_{x}k_{\perp}}{\gamma_{b}}=\frac{k_{\perp}^{2}c^{2}}{\omega^{2}}.
\end{equation}
The indices $p$ and $b$ denote the bulk plasma and the beam
respectively. The real part of the frequency
$\mathrm{Re}\omega\equiv\omega_{0}=k_{x}u_{x}+k_{\varphi}\upsilon_{\varphi}\approx
k_{x}u_{x}$ and the parameter $a$ is a complex number and the growth
rate of the instability $\Gamma=\mathrm{Im}a$ is the greatest when
\begin{equation}\label{}
    k_{\perp}^{2}\lesssim
    \frac{3\omega_{p}^{2}}{2\gamma_{p}^{3}c^{2}}.
\end{equation}
In this case one can write
\begin{equation}\label{}
    \Gamma\approx
    \left(\frac{n_{b}}{n_{p}}\right)^{1/2}\frac{\gamma_{p}^{3/2}}{\gamma_{b}^{1/2}}k_{x}u_{x}.
\end{equation}
The condition (13) implies the initial value for the wave vector
$k_{\perp}\sim1/r_{D}\sim10^{-1}$ (for the Crab pulsar parameters),
$r_{D}$ the Debye radius defines the scale of perturbations below
which the condition of quasi-neutrality is violated. Consequently,
this defines the maximal value for the perpendicular component of
the wave vector. On the other hand the minimal value for the wave
vector is confined by the perpendicular dimension of the
magnetosphere in the region of the wave generation $k_{min}\sim
1/R_{\perp}^{max}$. The perturbations with the frequency
$\omega_{dr}\approx k_{x}u_{x}$, which are propagating almost across
the magnetic field lines we will be calling the drift waves. The
drift waves are excited at the left slope of the distribution
function corresponding to the beam. The wave draws energy from the
longitudinal motion of the beam particles as in the case of an
ordinary Cherenkov wave-particle interaction. However, the wave is
excited only if $k_{x}u_{x}\neq0$, i.e., in the presence of drift
motion of the beam particles.

From expression (14) follows that the increment is greater the
farther from the pulsar surface the instability develops, while the
drift velocity strongly increases with distance $u_{x}\sim
r^{3}/R_{B}$. Generated in the light cylinder region the growth rate
of the drift waves estimated from expression (14) $\Gamma\sim
10\mathrm{s}^{-1}$ appears quite small that is natural as number of
the particles taking part in the resonance is not very big. However,
the drift waves propagate nearly transversely to the magnetic field,
encircling the magnetosphere, and stay in the resonance region for a
substantial period of time. Although the particles give a small
fraction of their energy to the waves and then leave the interaction
region, they are continuously replaced by the new particles entering
this region. The waves leave the resonance region considerably more
slowly than the particles. Hence, there is insufficient time for the
inverse action of the waves on the particles. The accumulation of
energy in the waves occurs without quasi-linear saturation.

Note that these low-frequency waves are nearly transverse, with the
electric vector being directed almost along the local magnetic
field. From Maxwell's equation
\begin{equation}\label{}
    \mathrm{rot }\mathbf{E}=-\frac{1}{c}\frac{\partial \mathbf{B}}{\partial
    t},
\end{equation}
after Fourier transformation one can obtain
\begin{equation}\label{}
    B_{r}=\frac{k_{x}c}{\omega}E_{\varphi}\sim E_{\varphi}\frac{c}{u_{x}}.
\end{equation}
Consequently for the drift waves $B_{r}\gg E_{\varphi}$.

For the nonzero parallel component of the wave vector, the resonant
condition (4) can be rewritten as:
\begin{equation}\label{}
    \omega=ku_{x}\sin\left(\frac{\pi}{2}-\theta^{\prime}\right)+
    k\upsilon_{\varphi}\cos\left(\frac{\pi}{2}-\theta^{\prime}\right),
\end{equation}
where $\theta^{\prime}=\pi/2-\theta\ll1$ taking into account that
$\theta\approx\pi/2$. This expression can be rewritten as,
$\omega\approx k(u_{x}+c\theta^{\prime})$ from which one can obtain
\begin{equation}\label{}
     \frac{\partial \omega}{ \partial k}\approx u_{x}+c
     \theta^{\prime}.
\end{equation}
For the condition $\theta^{\prime}\ll u_{x}/c$ only a small fraction
of the whole energy accumulated in drift waves is transferred in the
direction along the magnetic field lines in the low frequency range.
The energy is mainly transferred perpendicularly to the field lines
along the circular orbits. The time of energy pumping from the beam
electrons into the drift waves depends on the angle
$\theta^{\prime}=k_{\varphi}/k_{x}$. This process continues on the
linear stage of the turbulence until the wave does not leave the
magnetosphere. The time during that the beam particles leave the
pulsar magnetosphere is estimated as $\tau\sim
r_{LC}/c\sim1/\Omega$, here $r_{LC}=cP/2\pi$ is the radius of the
light cylinder and $\Omega$ is the rotation frequency and for the
Crab pulsar parameters $\tau\sim10^{-2}$s. Consequently, the time
during that the drift waves stay in the magnetosphere and can
accumulate the energy is of the order of $\tau(k_{x}/k_{\varphi})$.

At the linear stage of the resonance interaction of the energetic
beam electrons with the drift waves their drift velocity is quite
large. The particle move across the magnetic field with the velocity
of the order of $u_{x}\sim 10^{-2}c$, give part of their energy to
the waves and are carried out from the interaction region. New
particles enter this region, in their turn give part of their energy
to the waves and so on. The wave leaves the interaction region
considerably slower than the particles. Hence there is sufficient
time for the waves to accumulate energy. The energy accumulation
process is maintained until the nonlinear processes start to
operate. In the next section we consider the stage of development of
nonlinear processes and discuss the scenario of formation of GRPs.

\section{Nonlinear interaction of drift waves}

The energy of the excited drift waves grows when propagating across
the magnetic field and encircling the region of the open field lines
until the nonlinear process of induced scattering of waves on plasma
particles develops. From all nonlinear processes the probability of
the three-wave interaction and the induced scattering of waves on
plasma particles is the highest. However, the second-order current
describing the decay interaction is proportional to $e^{3}$. Hence
the contribution of electrons and positrons is compensated if their
distribution functions coincide. At the same time the induced
scattering is proportional to the even charge power ($e^{4}$). So
that the induced scattering on particles appears significant, it is
necessary that the number of particles taking part in the resonance
is sufficiently large, i.e. the bulk plasma particles with the
Lorentz factors $\gamma_{p}\sim2-5$ should be the resonant
particles. The drift velocity for the particles with small Lorentz
factors is considerably small and the particles move practically
along the field lines. Consequently, the nonlinear interaction
occurs along the pulsar magnetic field and the resonance condition
is written as \citep{ka_b}:
\begin{equation}\label{}
    \omega-\omega^{\prime}-(k_{\varphi}-k_{\varphi}^{\prime})\upsilon_{\varphi}=0.
\end{equation}
After taking into account that
$\omega=k_{x}u_{x_{b}}+k_{\varphi}\upsilon_{\varphi_{b}}$ and
$\upsilon_{\varphi_{p}}\approx c(1-1/(2\gamma_{p}^{2}))$, one can
rewrite the resonance condition (19) as
\begin{equation}\label{}
    \frac{k_{x}-k_{x}^{\prime}}{k_{\varphi}-k_{\varphi}^{\prime}}=
    -\frac{c}{u_{x_{b}}}\frac{1}{2\gamma_{p}^{2}}.
\end{equation}
This equation implies two cases: $(k_{x}-k_{x}^{\prime})>0$ and
$(k_{\varphi}-k_{\varphi}^{\prime})<0$, or
$(k_{x}-k_{x}^{\prime})<0$ and
$(k_{\varphi}-k_{\varphi}^{\prime})>0$.In the first case
$k_{x}^{\prime}<k_{x}$, the perpendicular component of the wave
vector is reducing via the scattering process and the parallel
component  grows $k_{\varphi}^{\prime}>k_{\varphi}$. For the second
case $k_{x}^{\prime}>k_{x}$, $k_{\varphi}^{\prime}>k_{\varphi}$.
Both of these processes are equiprobable. However, at the initial
stage of the development of scattering process the second case is
unlikely to develop. In Sec. 3 we have shown that for the linearly
generated drift waves propagating almost across the magnetic field
the wave vector, i.e. $k_{x}$ is maximal. The minimum possible value
for the wave vector is defined from the size of the magnetosphere
and the minimal value for the parallel component of the wave vector
for the Crab parameters $k_{\varphi}\sim1/r_{LC}\sim10^{-8}$cm.
Consequently, before the scattering takes place
$\theta^{\prime}\simeq k_{\varphi}^{0}/k_{x}^{0}\simeq 10^{-7}$ and
the time during that the waves stay in the pulsar magnetosphere
accumulating energy will be of the order of $\tau
(k_{x}^{0}/k_{\varphi}^{0})\simeq 10^{5}$s.

The scattering event causes growth of the parallel component of the
wave vector and the perpendicular component reduces at the same
time. At some point one can reach the case when $k_{x}^{\prime}\ll
k_{\varphi}^{\prime}$, in the other words the wave will be
propagating with the small inclination angle with respect to the
magnetic field lines. This must cause appearance of the radio
emission and supposedly in this case the GRP will be observed. The
time of escape of such waves from the pulsar magnetosphere is of the
order of $r_{LC}/c\sim 10^{-2}$s, as they are propagating almost
along the magnetic field (the angle $\theta\approx
k_{x}^{\prime}/k_{\varphi}^{\prime}\ll1$). The drift waves spend
much more time encircling the open field lines and accumulating
energy than escaping the magnetosphere to reach an observer.

\section{Application of the model and conclusions}

The GRPs are the brightest sources of radio emission from
astrophysical objects. They do not affect the average radio emission
characteristics of the given pulsar and are detected from pulsar
with parameters differing in wide range. The GRPs are distinguished
from pulsar's ordinary pulsed radio emission by several special
properties. The main property, as mentioned above is that the peak
intensities of GRPs greatly exceed the peak intensities of the
ordinary average pulses. For the Crab pulsar's strongest GRP, the
peak flux density exceeds the mean flux density of regular pulses by
a factor of $5\cdot10^{5}$ \citep{kosty}. Let us estimate the energy
gained by drift waves at the phase of perpendicular propagation. The
energy source of the process is the kinetic energy of the beam
particles with the typical Lorentz factors
$\gamma_{b}\sim10^{6}-10^{7}$ drifting in the perpendicular
direction of the inhomogeneous. Consequently, for the maximum
accumulated energy by the drift waves through their propagation in
the pulsar magnetosphere one can write
\begin{equation}\label{}
    mu_{x}^{2}\gamma_{b}n_{b}\tau\frac{k_{x}}{k_{\varphi}},
\end{equation}
here $\tau (k_{x}/k_{\varphi})$ is the accumulation time and
$mu_{x}^{2}\gamma_{b}n_{b}$ is the energy density of the beam
particles. In order to estimate the excess of the GRP peak intensity
over the ordinary pulse, one needs to consider the a different
generation mechanism. In particular, as we suggest the ordinary
radio emission is generated via the plasma instabilities.
Considering that the ordinary radio pulses in the Crab pulsar are
produced through the cyclotron-Cherenkov resonance, the mean pulse
peak energy can be estimated as
\begin{equation}\label{}
    mc^{2}\gamma_{b}n_{b}\tau.
\end{equation}
Here we have taken into account that the speed of the beam electrons
approximately equals $c$ and multiplied the energy density by the
escaping time of the particles from the magnetopshere $\tau$. Now
for the ration of the peak flux densities of the GRP and the mean
ordinary  pulse (OP) we obtain
\begin{equation}\label{}
    \frac{S_{GRP}}{S_{OP}}\simeq\left(\frac{u_{x}}{c}\right)^{2}\times\left(\frac{k_{x}}{k_{\varphi}}\right).
\end{equation}
For the Crab pulsar parameters $S_{GRP}/S_{OP}\sim10^{5}$ that
matches the observations.

The GRPs are very bright and short. As shown in \citet{hankins} the
GRPs from the Crab pulsar are as short as $2$ns. The resonant
particles are moving along the field lines to the observer and the
angular distribution of the radiation reaching the observer is
mainly concentrated within the small angle \citep{landau,schwinger}
\begin{equation}\label{}
    \alpha_{rad}\simeq\frac{1}{\gamma_{res}},
\end{equation}
around the field line. If the size of the emitting spot is
negligible (compared to the size of the magnetosphere) then the
duration of emission received by an observer is
\begin{equation}\label{}
    \tau_{rad}\simeq\frac{1}{\Omega\gamma_{res}}.
\end{equation}
Noting that for the crab pulsar $\Omega\approx200$s$^{-1}$ and for
the particles of primary beam duration of the pulse is
$\tau_{rad}\sim10^{-9}$s of the order of nanosecond.

As we can see two main characteristics, the excess of flux density
relative to an average pulse ones and a short pulse time-scale
compared to an average pulse is well explained in the framework of
the present scenario. The radio emission that we receive as giant
radio pulses is generated via the Cherenkov-drift resonance near the
light cylinder similarly as the ordinary radio pulses. The only
difference is the propagation direction of the generated waves. In
case of ordinary radio pulses the generated waves are propagating
almost along the magnetic field lines leaving the pulsar
magnetosphere and consequently the resonant region considerably
fast. On the other hand the same resonance provides generation of
almost perpendicular waves that encircle the region of open magnetic
field lines and propagate in the direction of the magnetic field
with the very low speed of the order of $ck_{x}/k_{\varphi}$. This
ensures presence of the waves in the resonant region for a
sufficiently long time. The resonant particles leave the
wave-particle interaction region quite fast, though the new
particles enter this region continuously and the waves appear to
have sufficient time to accumulate energy. The accumulated energy is
received by observer as pulsar radio emission due to change of the
direction of the mentioned perpendicular waves via nonlinear
scattering processes. As these type of waves have much more time to
gain particles' kinetic energy than the waves that directly leave
the magnetosphere, the pulses with very large energy excess are
observed. The observations reveal that for the Crab pulsar the GRPs
can occur anywhere within the average pulse, which could be caused
by excitement of the drift waves at different altitudes and magnetic
field lines with. This could also cause little changes in observed
frequency of the waves. One more fact in favor of our model is the
nondetection of correlation of GRPs and emission in higher
frequencies \citep{shea}. The mechanism of generation of high
frequency emission in pulsars differs from that of radio emission
(see \citet{ch}) and is connected with the appearance of pitch
angles due to development of cyclotron-Cherenkov and Cherenkov-drift
resonances, switching on the synchrotron emission generation
mechanism. The high frequency emission fulfills $\lambda\ll
n^{-1/3}$ condition leaving the generation region freely without
taking part in any process of energy accumulation by waves.

\section*{Acknowledgments}

This research was supported by Shota Rustaveli National Science
Foundation of Georgia (SRNSFG) [grant number FR/516/6-300/14]. Basic
Research Program of the Presidium of  the Russian Academy of
Sciences "Transitional and Explosive Processes in Astrophysics(P41)"
and Russian Foundation for Basic Research (grant 16-02-00954).

\bsp    
\label{lastpage}
\end{document}